\documentclass[pre,twocolumn,showpacs,floatfix,openany]{revtex4-1} 

\usepackage{graphicx}
\usepackage{dcolumn}
\usepackage{amsmath}
\usepackage{epsfig}
\usepackage{amssymb}
\usepackage{wrapfig}
\usepackage{sidecap}
\usepackage{xfrac}
\usepackage{footnote}
\usepackage{natbib}
\usepackage{float}
\usepackage{color}
\usepackage{lipsum}
\usepackage{footnote}

\newcommand{\beq}{\begin{equation}}
\newcommand{\eeq}{\end{equation}}
\newcommand{\bea}{\begin{eqnarray}}
\newcommand{\eea}{\end{eqnarray}}

\makeatletter
\def\btt#1{\texttt{\@backslashchar#1}}
\DeclareRobustCommand\bblash{\btt{\@backslashchar}}
\makeatother

\begin{document}

\title{Breakdown of Kinetic Compensation Effect in Physical Desorption}

\author{Nayeli Zuniga-Hansen}
\email{zunigahansen@lsu.edu}
\affiliation{Department of Physics \& Astronomy, Louisiana State University, Baton Rouge,
Louisiana, U.S.A.}

 \author{Leonardo E. Silbert} 
\affiliation{Department of Physics, Southern Illinois University Carbondale,
  Carbondale, Illinois 62901, U.S.A.}

\author{M. Mercedes Calbi} 
\affiliation{Department of Physics, University of Denver, Denver, Colorado
  80208, U.S.A.}

\date{\today}

\begin{abstract}

The kinetic compensation effect (KCE), observed in many fields of science,
is the systematic variation in the 
apparent magnitudes of 
the Arrhenius parameters $E_a$, the energy of activation,
and $\nu$, the preexponential 
factor, as a response to perturbations. 
If, in a series of closely related activated processes, these parameters
exhibit a strong linear correlation, it is expected that an isokinetic relation will occur, then
the rates $k$ become the same at a common compensation
temperature $T_c$.
The reality of these two phenomena continues to be debated as they have not been explicitly
demonstrated and their physical origins remain poorly understood.
Using kinetic Monte Carlo simulations on a model interface, we explore
how site and adsorbate interactions influence
the Arrhenius parameters during a typical desorption process.
We find that their transient variations result in a net partial compensation, due to the 
variations in the prefactor not being large enough to completely offset those in $E_a$, both in
plots that exhibit a high degree of linearity and in curved non-Arrhenius plots.
In addition, the observed isokinetic relation arises due to a transition
to a non-interacting regime, and not due to compensation between $E_a$ and $\ln{\nu}$.
We expect our results to provide a
deeper insight into the microscopic events that 
originate compensation effects and isokinetic relations in our system,
and in other fields where these effects have been reported.

\end{abstract}

\pacs{PACS numbers: 82.20.Db, 68.43.Vx, 68.43.Nr, 68.43.De}

\maketitle



\section{Introduction}
Many physical, biological, and chemical processes exhibit a strong temperature
dependence, in the sense that they rely on thermally activated mechanisms to
overcome energy barriers in order for the process to proceed.
The rate, $k$, of many such processes follows an Arrhenius
type behavior:
\beq 
k = \nu e^{\frac{- E_a}{k_B T}},
\label{eq:Arrhenius}
\eeq 
where $E_{a}$ is the activation energy, $\nu$ the preexponential
factor, $T$ is the temperature
and $k_{B}$ is Boltzmann's constant. Information parameterizing such processes are usually obtained through an
Arrhenius plot, constructed with a series of measurements of
$\ln{k}$ vs. $\sfrac{1}{T}$, from which one obtains the activation energy as
the slope, and the preexponential factor from the intercept.

\begin{figure}
     \includegraphics[width=0.25\textwidth]{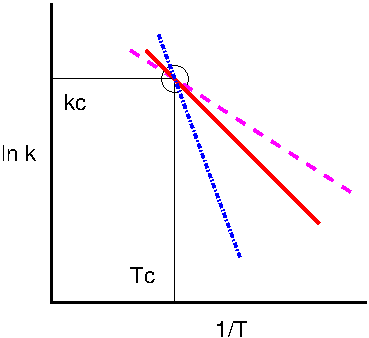}
     \caption{(Color Online) Schematic idealized Arrhenius plots for a set of
       closely related thermally activated processes: $\ln k = \ln \nu -
       \sfrac{E_{a}}{k_{B}T}$, crossing at the compensation temperature,
       $T_{c}$, where the rates appear to have the same value $k_{c}$. 
       When the parameters $E_a$ and $\nu$ are constant, the
       slope yields the energy of activation and the $y-$intercept yields the
       natural logarithm of the preexponential factor.}
\label{fig:sticker}
\end{figure}

\begin{figure}
     \includegraphics[width=0.4\textwidth]{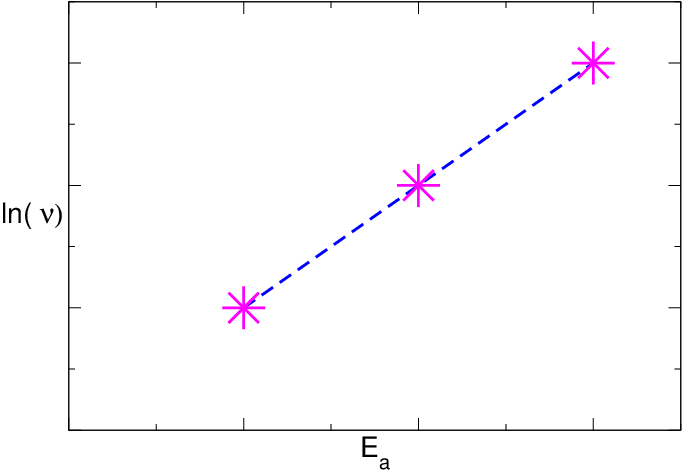}
     \caption{(Color Online) Schematic idealized Constable plot for a set of
       closely related thermally activated processes. Each data point represents 
	a data pair ($E_{a,i}$,$\ln{\nu_i}$) extracted from the slope and $y-$intercept,
	respectively, of the Arrhenius plot that corresponds to the $i^{th}$ activated
	process. The fingerprint of the kinetic compensation effect is said to be a strong 
            linear correlation among data points in this plot.}
\label{fig:Constable}
\end{figure}

A characteristic feature in a series of closely related thermally activated
processes, where a parameter has been varied (e.g. the concentration
of an additive in a chemical reaction)
is a systematic change in the apparent magnitudes of $E_a$ and $\nu$
\cite{LiuGuo:01,Lvov:13,Yelon:12} as a response to perturbations, known as the
\textit{kinetic compensation effect}. Such a hypothetical set is schematically represented in 
Fig.~\ref{fig:sticker}.
The premise behind the concept
of `compensation' is that a change in the magnitude of $E_a$ is compensated or
offset by a concomitant change in the preexponential factor 
$\nu$ \cite{Yelon:12,Estrup:86,Kreuzer:88,Tomkova:98,Lvov:13}, and they
satisfy the linear relationship \cite{Freed:11,Kreuzer:88}:
\beq 
\ln{\nu_i} = \beta  E_{a,i} + \alpha,
\label{eq:KCE}
\eeq
where $\alpha$ and 
$\beta$ are constants. This linear relationship originates
from the Constable plot, which is, in turn, contructed with the data pairs $E_{a,i},\ln{\nu_i}$ obtained from the
slope and $y-$intercept, respectively, of the $i^{th}$ Arrhenius plot \cite{Barrie:12b,Gottstein:98,LiuGuo:01}.
A hypothetical Constable plot of $\ln{\nu}$ vs. $E_a$ is schematically
represented in Fig.~\ref{fig:Constable}.

Following this method of analysis, a kinetic compensation effect is observed when the 
data points on the Constable plot fall on a straight line \cite{Barrie:12a}. Then
the parameters are obtained by combining Eq.~\ref{eq:KCE} with Eq.~\ref{eq:Arrhenius}:
$ln{\nu_i} = \ln{k_i} + \frac{E_{a,i}}{k_B T} = \beta E_{a,i} + \alpha$; 
$\beta$ then becomes $\sfrac{1}{k_B T_c}$, where $T_c$ is the \textit{compensation temperature}
\cite{LiuGuo:01,Barrie:12b,Yelon:12,Freed:11}, and $\alpha$ becomes $\ln{k_i}$.
This predicts that at $T_c$ the Arrhenius plots cross, and at that point
the rates become the same and
\textit{`independent of external parameters
and perturbations'} \cite{Barrie:12a,Douglas:09,Freed:11,LiuGuo:01,Linert:89}. 
This is known as the 
isokinetic relation (IKR) \cite{Barrie:12a,LiuGuo:01}, or isokinetic equilibrium, 
also schematically represented in Fig.~\ref{fig:sticker}.
Historically, the KCE and IKR have been defined interchangeably or as synonymous \cite{Agrawal:86}, 
since the
observation of one is thought to directly imply the occurrence of the other
\cite{LiuGuo:01,Agrawal:86}. This is a natural assumption, given that, as seen before, one can predict 
the IKR from a the linear relationship in Eq.~\ref{eq:KCE}. We should note, however,
that this is only possible when the linear correlation coefficient
between data points in the Constable plot is exactly $1$ \cite{LiuGuo:01}.

In their extensive review \cite{LiuGuo:01}, Liu and Guo propose that the KCE and IKR are
separate phenomena, which may be observed independently, and should be characterized as such:
the KCE should be identified solely by the strong linear correlation between $E_a$ and $\ln{\nu}$ 
\cite{Kreuzer:88,LiuGuo:01, Barrie:12a,Barrie:12b, Zhdanov:91,Linert:89}, and 
the IKR by the convergence of Arrhenius plots around
a single temperature value \cite{LiuGuo:01,Linert:89}.

The extraction of parameters from the slope and $y-$intercept of an 
Arrhenius plot has proven to be a useful and important method that
allows for the empirical determination of rates \cite{Barrie:11,Agrawal:86},
however, it has been widely accepted for some time now
that the parameters $E_a$ and $\nu$ need not be constants 
throughout many activated processes
\cite{Fichthorn:91,Fichthorn:94,Kreuzer:88,Kreuzer:89,
Zhdanov:89,Zhdanov:81,Kang:90}.
This leads to the observation that each point in the Constable plot indeed corresponds
to a data pair of \textit{fixed} values of
$E_{a,i}, \ln{\nu_i}$, therefore, if the two phenomena are to be defined separately,
this raises the question of the physicochemical significance of
an observed linear correlation in a case where the IKR is not simultaneously satisfied. 

It has been proposed that throughout an activated
process the parameters vary in a manner such that they increase or decrease in the same direction,
and \textit{compensate or offset} each other, in a way that the overall rate remains 
almost unchanged \cite{Kang:90,Fichthorn:94,Gottstein:98,Estrup:86,
LiuGuo:01,Lvov:13,Douglas:09,Tomkova:98,Yelon:12,Zhdanov:81,Dunitz:95,Kreuzer:88}.
This behavior would perhaps justify the extraction of constant Arrhenius parameters, but 
has not been explicitly
demonstrated\cite{Miller:87,Nieskens:03,LiuGuo:01}.
A strong interdependence of the parameters has been attributed
to them being extracted from the same temperature dependent
data \cite{LiuGuo:01,Barrie:12a,Cornish:02}, instead of through 
independent measurements 
(which are not 
always possible). A strong linear correlation obtained this way
is likely, as mentioned by Cornish-Bowden in \cite{Cornish:02}, 
the consequence of the two variables being `larguely
the same variable looked at in two ways'.
In addition, some instances of the IKR do not yield a
compensation temperature $T_c$ that 
falls within the experimental range, and is therefore found by 
extrapolation \cite{Yelon:12, Cornish:02}. 
Thus the existence of the kinetic compensation effect and isokinetic relation continue to be the
subject of heated debate, and
are often believed to have a purely mathematical origin
\cite{Koga:91,Barrie:11,Barrie:12a,Barrie:12b,Cornish:02} and to lack any physical or chemical
significance.

Nevetheless, the KCE and IKR, as well as the closely related 
entropy-enthalpy compensation, 
continue to be reported in many different areas of science, such as
temperature programmed desorption \cite{Miller:87}, fouling \cite{Barrie:11},
grain boundary migration \cite{Gottstein:98}, heterogeneous catalysis
\cite{Lvov:13}, crystallization of amorphous solids \cite{Banik:13}, glass
transitions \cite{Dyre:86}, adsorption \cite{Rawatt:09,Migone:13, Freed:11},
chemical reactions \cite{Freed:11} molecular self-assembly
\cite{Douglas:09,Kim:13} and the melting of solids \cite{Dunitz:95} among
others.

Here, we explore the very notion of the KCE, at the fundamental level, 
through the use of kinetic Monte
Carlo \cite{Voter:book} simulations in the context of physical desorption;
to do this we quantify the transient variations in the Arrhenius parameters
throughout the activated process and verify if those changes exhibit a compensatory behavior.
Our methodology is to perform the \textit{ab initio} numerical calculation of the 
energy of activation, along with the calculation of
the decrease in the coverage with increasing temperature,
during the thermal desorption of
interacting and non-interacting adsorbates from an
energetically homogeneous, crystalline lattice. 
By doing this we step aside from an approach that
relies on preconceived functional forms that yield the best fit to the
desorption data \cite{Tomkova:98,Barrie:12a}, and/or that predicts the variations in the parameters
to yield a KCE and an IKR, in the context of the criteria used to identify them in \cite{LiuGuo:01}.
In the present work 
we consider only attractive lateral interactions.
Using our numerical results for $E_a$ and coverage, we can
extract the preexponential factor $\nu$, and quantify
its transient variations as well. To our knowledge, this approach has not been done before.

\begin{figure}
  \includegraphics[width=1.1in]{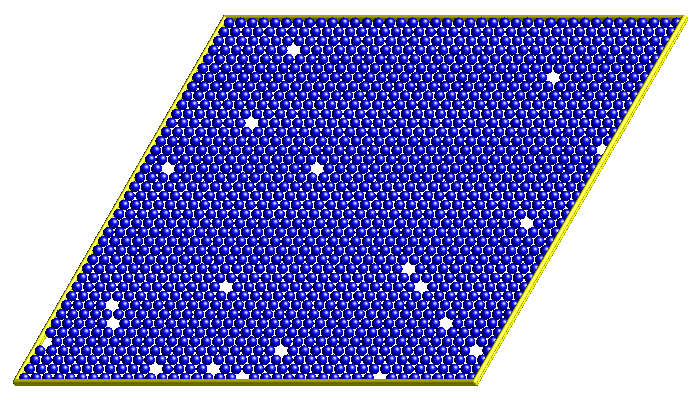}
  \includegraphics[width=1.1in]{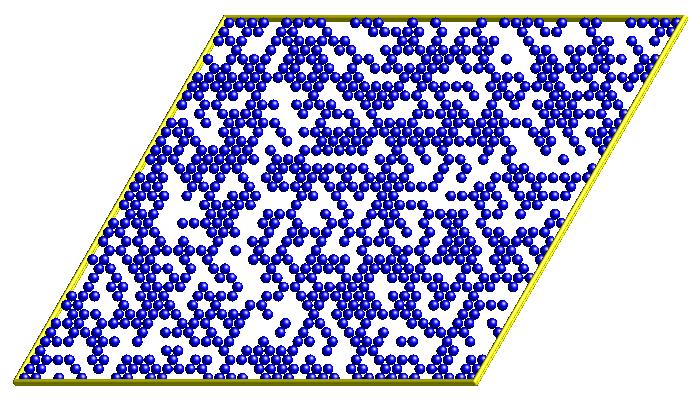}
  \includegraphics[width=1.1in]{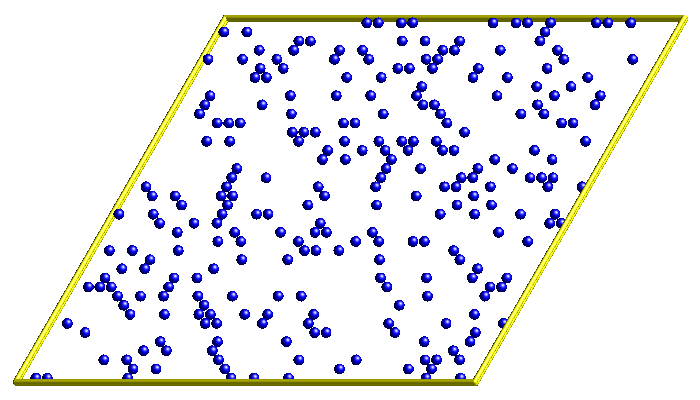}
  \caption{(Color Online) Visualization of the kinetic Monte Carlo simulations
    during a desorption run for a 2D square lattice substrate. Increasing
    temperature from left to right. In these instantaneous snapshots the
    (blue) filled circles represent occupied sites and the lines denote the
    periodic boundaries of the simulation domain. Binding energy is set to
    $E_{b}=100$ and the interaction energy $\epsilon=0$, in this
    representative example.}
\label{fig:snapshots}
\end{figure}

Our numerical results span a range of adsorbate-adsorbate attractive interaction strengths,
calculated as a percentage of the fixed surface binding energy, this provides the 
experimental parameter that is being altered in the series of similar activated processes; all
while keeping the substrate structure fixed.  
Our method allows us to verify and
explicitly quantify the level of
compensation that has not been successfully achieved to date using more
traditional methods.
Snapshots of
our computer simulations are shown in Fig.~\ref{fig:snapshots}. 

Our results show that the parameters effectively exhibit a behavior that supports the occurrence
a compensation effect, however, the observed changes in
the preexponential factor \textit{are not large enough to effectively compensate or offset
the significant variations in the activation energy, $E_a$},
which arise due to strong coverage dependence. This produces a net \textit{partial} compensation
between $E_a$ and $\ln{\nu}$ in all regimes of interaction strength presented
in this study. 
These results are in stark contrast with previously reported
strong coverage dependencies of the preexponential factor in the regime of strong interactions 
\cite{Fichthorn:94,Zhdanov:81,Zhdanov:89},
which can result when the parameters are extracted through forced linearization \cite{Nieskens:03,
Miller:87,Fichthorn:94}.
Partial compensation has been previously considered \cite{LiuGuo:01,Yelon:12}, but
this notion is not as widespread as that of a complete KCE;
here we also observe that the
partial compensation effect is not incompatible with 
the occurrence of curved Arrhenius plots, also
referred to as non-Arrhenius behavior \cite{Kang:90, Kreuzer:89,Limbach:06,Rossetti:13}.

In the context of isokinetic equilibrium, we observe a tendency of the 
Arrhenius plots to converge towards the region of low coverage and high temperature.
Our plots, however did not cross at a common $T_c$, instead the interacting 
regime plots reach the non-interacting one at
significantly different temperatures, which could be due to the $10\%$ difference between 
regimes of interaction strength.
Nevertheless, we extracted those temperatures,
as well as the values of $E_a$ and $\nu$ at those points,
and found that the Arrhenius parameters
become numerically very close to those assigned to the bare surface,
which implies that the convergence of rates cannot be attributed to 
a compensation effect between $E_a$ and $\nu$, but instead to a transition
to a regime where the effects of lateral interactions become negligible.
And it is also consistent with the rates 
becoming independent of perturbing parameters.

We also analyzed a subset of data with
adsorbate-adsorbate interaction strenght of $<10\%$ of the binding energy of the surface. 
Many physisorption experiments
with weakly interacting adsorbates and energetically heterogenous, crystalline surfaces fall in this
category \cite{Ulbricht:02}. We performed the linear fit to the Arrhenius plots
and show that the slope and $y-$intercept method of analysis yields more accurate results in this regime,
and also that, in this regime, the assumption of a constant preexponential factor can be justified.
Nevertheless, the parameters exhibit the same behavior we observed in the other stronger interaction
regimes, as well as the tendency towards an IKR, which can
also be explained in terms of a transition to the non-interacting regime.

\section{Model system and Methodology}

\begin{figure}
\begin{center}
     \includegraphics[width=3in]{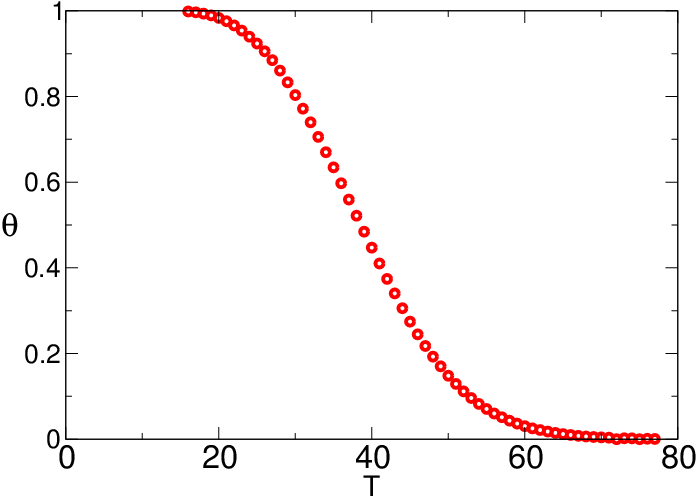}
     \caption{(Color Online) Typical desorption profile for a numerical TPD
       experiment of non-interacting adsorbates for a 2D square lattice
       substrate starting from full fractional coverage, $\theta = 1$. As the
       temperature, $T$, increases the coverage, $\theta$, decreases. The site
       binding energy is set to $E_{b}=100$ and the interaction energy
       $\epsilon=0$ (in simulation units), in this representative example.}
\label{fig:coverage}
\end{center}
\end{figure}

\begin{figure}[t!]
\begin{center}
    \includegraphics[width=1.67in]{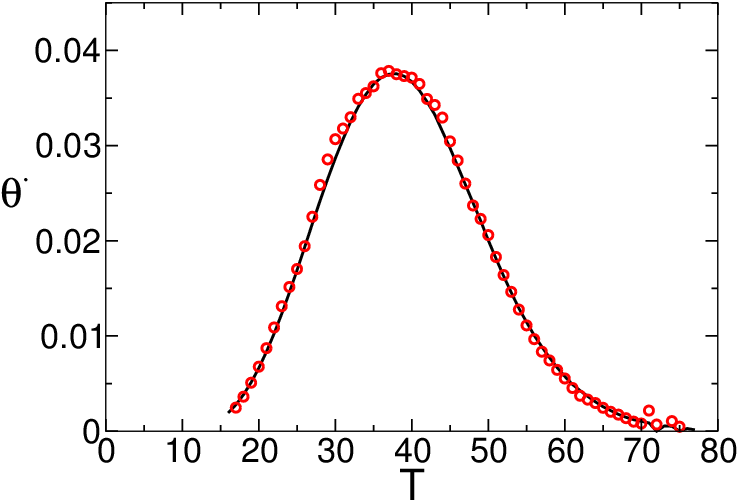}
    \includegraphics[width=1.67in]{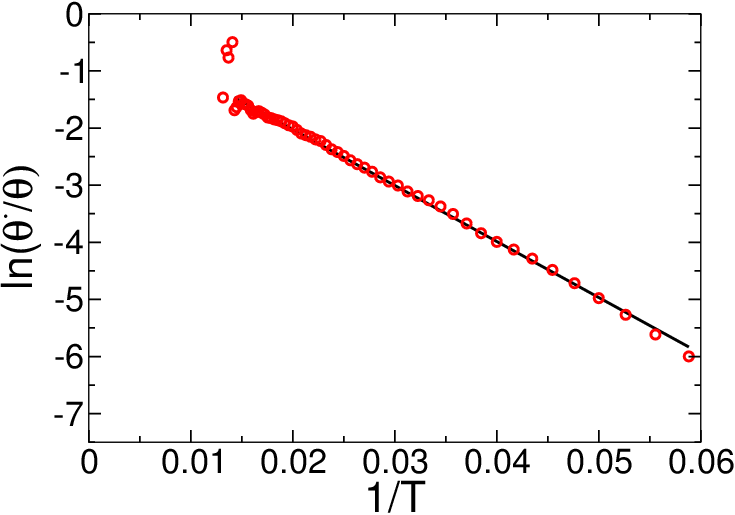}
    \caption{(Color Online) Rate of coverage decrease $\dot{\theta}$, as a
      function of temperature $T$ (left panel) obtained from the numerical
      derivative of the desorption data (Fig.~\ref{fig:coverage}) and the
      corresponding Arrhenius plot (right panel) for the thermal desorption of
      non-interacting species ($\epsilon = 0$) from a square lattice with
      homogeneous binding energy $E_{b} = 100$. The solid lines represent the
      fit to Eq.~\ref{eq:PolanyiW} on the left panel, and to the Arrhenius
      plot, $\ln{k} = -\beta E_a + \ln{\nu}$, with $k \equiv
      \theta/\dot{\theta}$, on the right panel. Note that scatter in the
      numerical data becomes more prevalent at higher temperatures.}
\label{fig:Noint}
\end{center}
\end{figure}

TPD is an experimental technique used in surface science to extract surface
parameters, such as binding energies, sample porosity \cite{Hansen:98} and
sorption capacity; it has applications in chemical speciation
\cite{Coufalik:14} and contaminant removal \cite{Smith:01}. In a typical
experiment a surface in an evacuated chamber is exposed to a gas until the
desired uptake is achieved, then the sample is heated with a linear
temperature ramp of the form: $T(t) = T(0) + \Delta T t$, where $\Delta T$ is
the temperature step and $t$ is time. The results are in the form of the
substrate fractional coverage, $\theta(T)$, as a function of
temperature $T$ \cite{Redhead:62}. A typical simulation data set is shown in
Fig.~\ref{fig:coverage} for the non-interacting species examples shown in
Fig.~\ref{fig:snapshots}. The corresponding rate and Arrhenius plots are
shown in in Fig.~\ref{fig:Noint}.

The most common method of analysis starts with the Polanyi-Wigner equation of
desorption:
\beq 
\dot{\theta} = \theta^n \nu e^{-E_a/k_BT},
\label{eq:PolanyiW}
\eeq 
where, $\dot{\theta}$ is the rate at which the coverage decreases with
increasing temperature, and $n$ the order of the process. Physical desorption
from a uniform planar surface corresponds to order $1$, thus we set $n=1$ for
the remainder of the present work.  The most common method of extraction of
the parameters of interest is from the slope and $y-$intercept, respectively, of the Arrhenius plot,
$\ln{k}$ vs. $\sfrac{1}{T}$, where in our case, $k \equiv \dot{\theta}/\theta$
(see Fig.~\ref{fig:Noint}).  As discussed before, such a parameterization has proven useful in the
empirical determination of rates \cite{Barrie:11}, however, the parameters $E_a$ and $\nu$ exhibit 
variations throughout the desorption process due to
one or more of the following factors: surface energetic heterogeneity
\cite{Burde:09,Zuniga:12a}, lateral interactions
\cite{Miller:87,Nieskens:03}, multiple chemical species \cite{Burde:10}, 
and/or changes in surface configuration \cite{Estrup:86}.

Our methodology is to simulate a TPD process from a quasi-two dimensional,
square lattice of side $L$, with $N=L^{2}=1600$ sites and periodic boundary
conditions using a kinetic Monte Carlo algorithm \cite{Voter:book}.
The lattice is energetically homogeneous, so that each site $j$, has an
associated binding energy, $E_{jb}=E_{b}=100$, in units where $k_{B}=1$. We
explore attractive interaction strengths $\epsilon$, that range from $0$ to 
$\epsilon \leq 0.9 E_{b}$. To track the desorption process, the kinetic Monte Carlo
scheme follows a series of steps: we first specify the initial conditions,
including the binding and interaction energies, initial temperature (which we
alter depending on $\epsilon$), step size, and initial coverage (which is set
to $100\%$ in all cases). The second step is to calculate the number of
occupied nearest neighbors per site, the site energies, and the probabilities
associated with each of the allowed transitions. The energy 
desorption barrier per site $E_{j}$, is given
by
\beq 
E_{j} = E_{b} + \sum_{m=1}^{z} n_{jm} \epsilon,
\label{eq:Esite}
\eeq 
where each site, $j$, picks up an energy contribution from its $m$ nearest
occupied neighbors. Thus, $n_{jm} = 1$ when a neighbor site is
occupied, zero if empty, and $z=4$ is the coordination number for the square
lattice under consideration here. Next, an allowable transition - desorption
or diffusion to a neighboring site - is selected and the state of the system
is updated, then the time counter is increased as prescribed by the kinetic
Monte Carlo algorithm within the grand canonical ensemble.
The temperature ramp is controlled by increasing $T$ after so many time steps
$\delta t$, which is here set to unity. Finally the coverage, temperature and
activation energy are updated and recorded. The process is repeated until the lattice is completely
empty. Our results are obtained as an (ensemble) average over 100, independent
runs; see Fig.~\ref{fig:snapshots} for representative simulation images at
early (low), intermediate, and late (high) times (temperatures).

\section{Results}
%
%
The first step is to verify that our simulation results can be fitted to Eq.~\ref{eq:PolanyiW}
for the non interacting regime ($\epsilon = 0$).We show this on the left panel of Fig.~\ref{fig:Noint}.
On the right panel of the same figure, we performed the corresponding 
linear fit to the Arrhenius plot, from which we obtain the following parameters: $E_a = 100 \pm
2$  and $\nu=1 \pm 0.01$ (extrapolated intercept). The value for $E_{a}$, as expected, matches
the input binding energy, $E_{b}$, within error. 
\begin{figure}
  \includegraphics[width=0.5\textwidth]{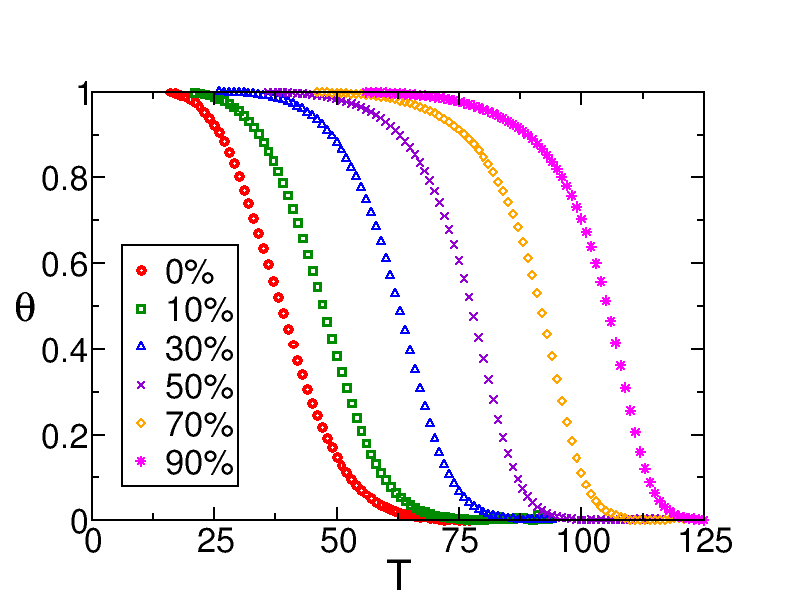}
  \includegraphics[width=0.5\textwidth]{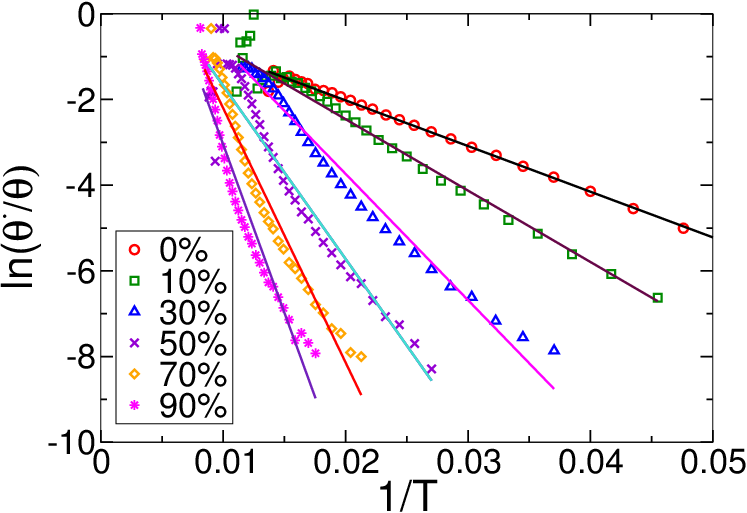}
  \caption{(Color Online) Desorption data for different interaction energies
    $\epsilon$ specified as a percentage of the fixed binding energy, $E_{b} =
    100$. Desorption curves (top) shows the dependence on the fractional
    coverage $\theta$, as a function of temperature $T$. Arrhenius plots
    (bottom) $\ln{\sfrac{\dot{\theta}}{\theta}}$ vs.$ \sfrac{1}{T}$ and best
    linear fits to the data (lines). As the interaction strength increases, so
    does the curvature of the plot.}
\label{fig:Aplots}
\end{figure}

\subsection{Activation Energy}
One of our main results is the calculation of the
transient variations in the energy
of activation throughout the desorption process, during which
$E_a$ is calculated using Eq.~\ref{eq:Esite} as, 
$E_{a} = \frac{1}{N}\sum_{j}^{N}E_{j}$.

\begin{figure}[h]
    \includegraphics[width=0.45\textwidth]{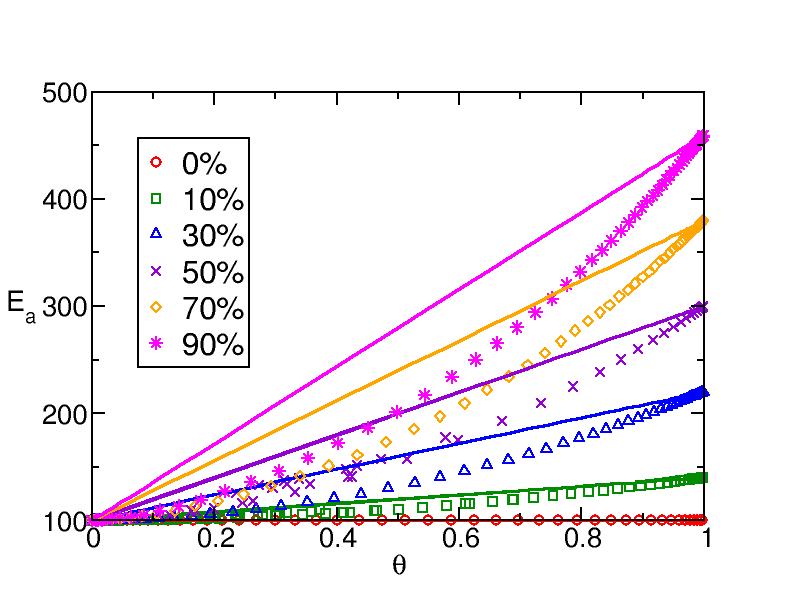}
    \caption{(Color Online) Magnitude of the activation energy $E_{a}$, as a
      function of coverage $\theta$ (symbols). Solid lines are a mean field
      (MF) analysis of the data: $<E_{a}> = (1+zf\theta)E_{b}$, with
      $f=\epsilon/E_{b}$ (corresponding to the percentage values given in the
      legend) and $z$ is the coordination number of the lattice. Here, $z=4$
      for the square lattice.}
\label{fig:Ea}
\end{figure}

Our results for $E_a$ are plotted as a function of coverage in Fig.~\ref{fig:Ea}. In the
non-interacting regime, the activation energy remains constant and matches the
binding energy itself, as expected, which is consistent with the fact that it represents
the only energy barrier to desorption. This feature of the non-interacting
regime also applies locally at each site. In the case of interacting species,
on the other hand, additional contributions coming from site-occupied nearest
neighbors result in a stronger binding of the adsorbates to the surface, which
varies locally due to the heterogeneous distribution of interacting occupied
sites throughout the desorption process. This leads to an enhanced activation
energy, or effective desorption barrier, and also to the curvature of Arrhenius plots,
which is observed to increase
as a function of $\epsilon$ for attractive interactions.

\begin{figure}[h]
    \includegraphics[width=0.45\textwidth]{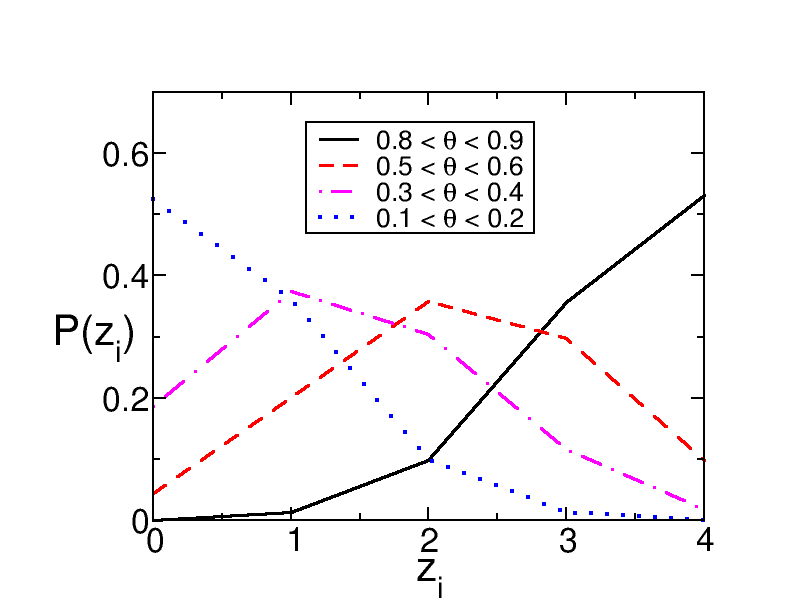}
    \caption{(Color Online) Distributions, $P(z_{o})$, of the interacting
      site-occupied coordination number $z_{o}$, at several values of the
      coverage,$theta$ indicated by key, for the representative example,
      $\epsilon = 30\%$ of $E_{b}$. Other non-zero interaction energies
      exhibit similar behavior.}
\label{fig:z}
\end{figure}

In a MF way, the additional contributions to the activation energy due to
occupied sites interacting with occupied neighbors is given by the term:
$z\theta\epsilon$, where $z$ is the coordination of the lattice substrate ($z=
4$ for the square lattice used here), $\theta$ is the coverage, and $\epsilon$
the adsorbate-adsorbate interaction energy. Then, the total mean activation
energy is, $<E_{a}> = (1+zf\theta)E_{b}$, where $f = \epsilon/E_{b}$, is the
fractional interaction energy. This analysis is shown by the solid line fits
to the data in Fig.~\ref{fig:Ea}. As seen, the MF approach is only correct for
the non-interacting regime (when $f=0$) and for the interacting systems only
at the extreme coverage values (when $\theta=1$ and $0$). The reason behind
this is that the MF approach presupposes that at the molecular level, each
site sees the same number of occupied neighbors throughout the substrate; in
other words the coordination of occupied interacting sites, $z_{o}$, is
delta-function distributed. This is only true at complete coverage, where the
distribution of interacting sites, $P(z_{o}) = \delta(z_{o}-4)$, and again at
zero coverage: $P(z_{o}) = \delta(0)$. Thus, deviations from the MF picture
come from the distribution of interacting sites during the kinetics. As a
result, due to the added energy contributions when interactions are present,
the sites become energetically heterogeneous. To illustrate this point,
Fig.~\ref{fig:z} shows the $P(z_{o})$ for one interaction energy at several
coverages, which shows how the occupation coordination becomes wider at
coverages away from complete and zero coverage.

\subsection{Pre-exponential Factor}
\begin{figure}
    \includegraphics[width=0.45\textwidth]{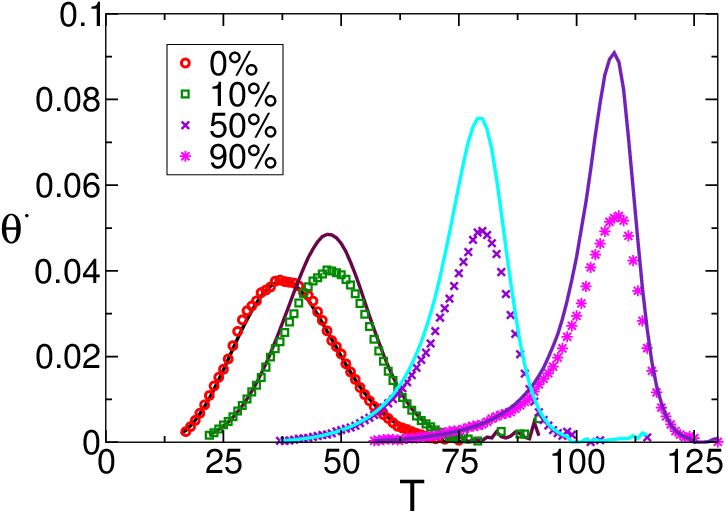}
    \caption{(Color Online) (Color Online) Rate of coverage decrease,
      $\dot{\theta}$, computed as the numerical time-derivative of the
      coverage data. The solid lines compare the rates calculated using the
      activation energies and Eq.~\ref{eq:PolanyiW} assuming $\nu = 1$. The
      difference with increasing interaction is a signature of deviations from
      Arrhenius behavior due to changing $\nu$.}
\label{fig:EaPW}
\end{figure}
\begin{figure}
    \includegraphics[width=0.45\textwidth]{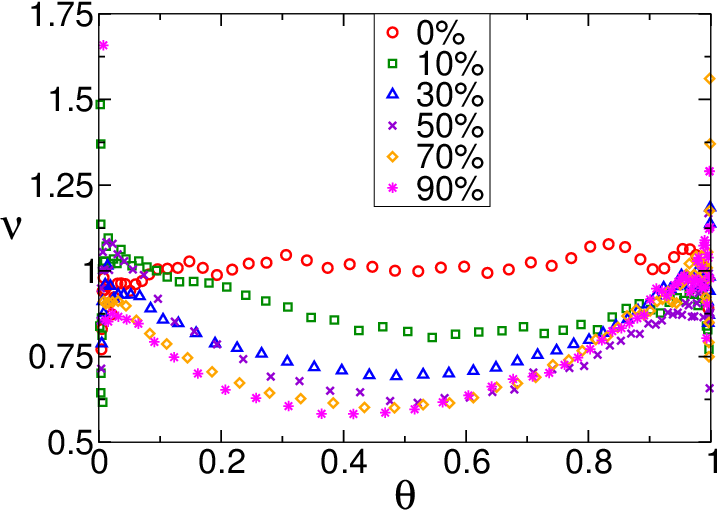}
    \caption{(Color Online) Preexponential factor $\nu$, as a function of
      coverage $\theta$, for different interaction energies.}
  \label{fig:nuvscov}
\end{figure}

To verify whether a compensation effect occurs, we calculated
the rate of desorption analytically, using the energies of activation in
Fig.~\ref{fig:Ea} and Eq.~\ref{eq:PolanyiW}, while keeping $\nu$ fixed at $1$.
These results are plotted in Fig.~\ref{fig:EaPW} (solid lines).
We then compared the resulting plots with the desorption rate data, obtained through 
the numerical derivatives of
the desorption curves on the top panel of Fig.~\ref{fig:Aplots},
and are also plotted in Fig.~\ref{fig:EaPW} (symbols).
The difference between the two indicates that there must
be some variation in the prefactor $\nu$.
We then extracted $\nu$, these results
are plotted in Fig.~\ref{fig:nuvscov}, where it can be seen that the preexponential
factor also exhibits variations as the coverage decreases.
In the non-interacting regime, Fig.~\ref{fig:nuvscov} displays how $\nu$
remains constant at unity (the observed fluctuations are due to the numerical
derivatives used to obtain the data). For $\epsilon > 0$, $\nu$ exhibits a
systematic deviation from the non-interacting value as the strength of the
interaction increases. In this sense $\nu(\epsilon=0)$ is the bare desorption
rate that is renormalized in the presence of interactions.  While the trend
towards decreasing $\nu$ with increasing $\epsilon$, is consistent with some
level of compensation, the changes in $\nu$ are significantly smaller in
magnitude compared to those in $E_a$ (see Fig.~\ref{fig:Ea}), and are not
large enough to effect complete compensation.

\begin{figure}
    \includegraphics[width=1.67in]{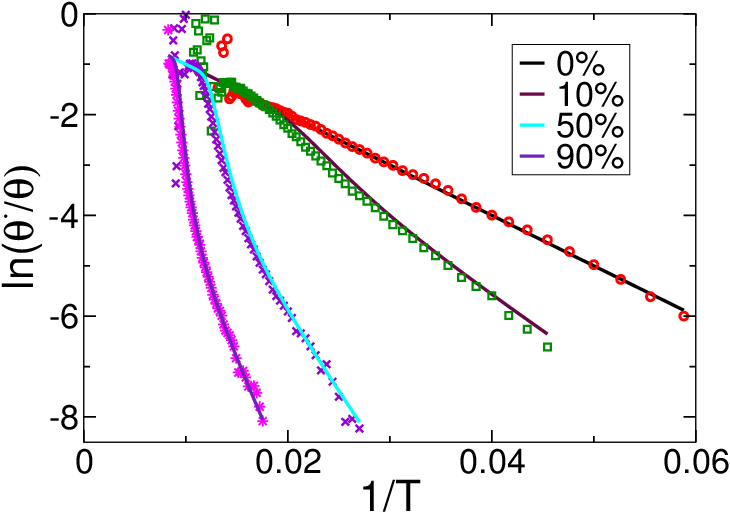}
       \includegraphics[width=1.67in]{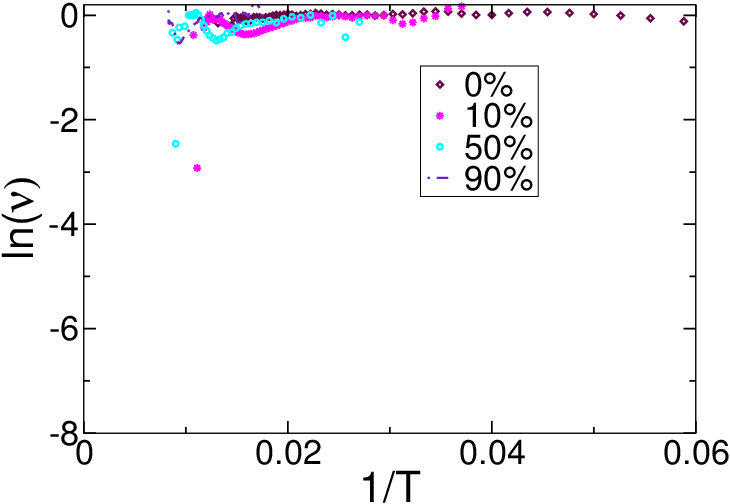}
       \caption{(Color Online) Arrhenius plots with the contribution from
         $E_a$ (solid lines), $\nu$ is kept fixed at $1$ (left panel) compared
         to the Arrhenius plots obtained from our data (symbols); and the
         natural logarithm contribution to the Arrhenius plots for these
         interaction regimes (right panel) using the same scale as the figure
         on the left panel for direct comparison.  Most of the curvature in
         the Arrhenius plots is due to the variations in $E_a$, which shows
         that the contribution from $\nu$ does not compensate for the changes
         in $E_a$, as variations in $\nu$ are small by comparison, as can be
         seen in the right panel. Scatter is due to numerical noise.}
\label{fig:AplotsEnu}
\end{figure}

To visualize the degree to which the parameters compensate each other, we
plotted the separate contributions of $E_a$ and $\ln{\nu}$ to the Arrhenius curve of 
various interaction strength regimes, these results are shown in Fig.~\ref{fig:AplotsEnu}.
On the left panel we show the Arrhenius plots calculated by keeping $\nu$
fixed at a value of $1$, and using our numerical results for $E_a$ (solid lines).  When compared to
the Arrhenius plots obtained directly from the desorption data
(symbols), the curvature remains almost unchanged, indicating that the relative
contribution due to $\ln{\nu}$ is small in all regimes of interaction strength.
On the right panel of Fig.~\ref{fig:AplotsEnu} 
we added the
we plot $\ln \nu$ vs. $\sfrac{1}{T}$, and 
we see that $\ln \nu$ remains
approximately constant and relatively close to $1$ in all cases. 
Our results in Figs.~\ref{fig:nuvscov} and
Fig.~\ref{fig:AplotsEnu} do not yield `unusually
large' preexponential factors that exhibit a
strong coverage dependence in the regime of strong
interactions \cite{Zhdanov:81,Zhdanov:89,Seebauer:88,Pfnur:78},
and which have been
identified as an indicator of false compensation effects in thermal desorption
\cite{Nieskens:03}.

\subsection{Correlations}
\begin{figure}
    \includegraphics[width=1.68in]{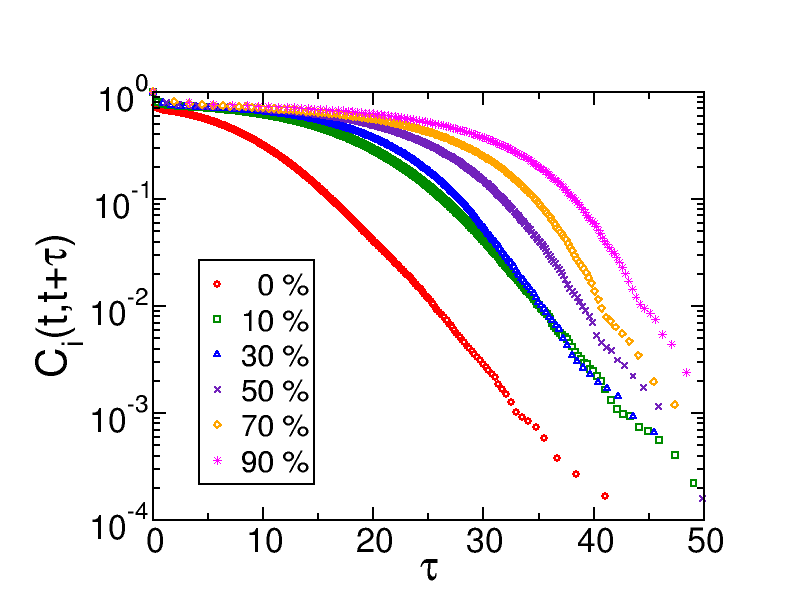}
    \includegraphics[width=1.68in]{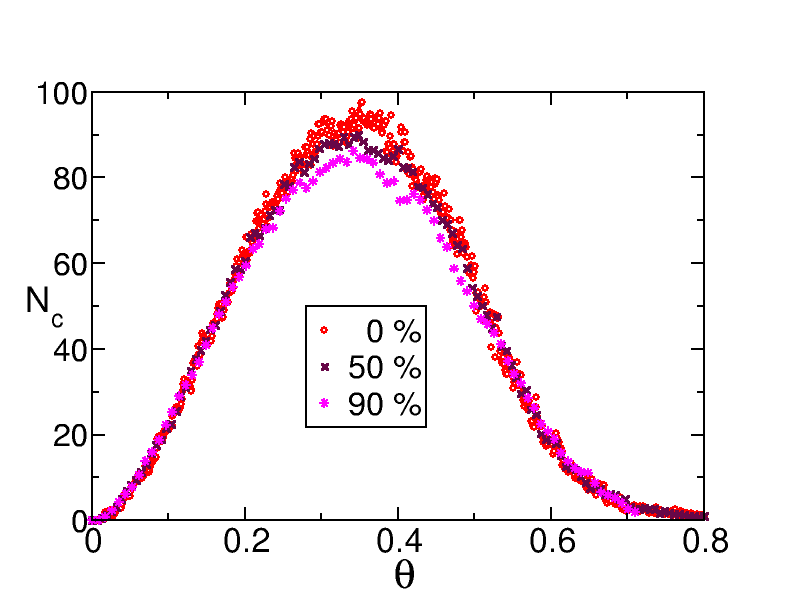}
    \caption{(Color Online) Correlation and cluster analyses. Left: site time
      autocorrelation function $C_{j}$ (see text for definition), for various
      interaction strengths. Simulation times were chosen to span the
      desorption processes for the different energies. Right: number of
      distinct clusters, $N_{c}$, as a function of coverage for zero,
      intermediate, and large interaction strengths.}
  \label{fig:Cit}
\end{figure}

A thermodynamic point of view posits that the changes in $\nu$ can be
attributed to changes in the entropy \cite{Sharp:01,Gottstein:98}. This view
is somewhat contained within the Erying-Polanyi equation
\cite{Erying:35,Erying:35a,Yelon:12}, where $E_a$ is associated with the enthalpy of
activation $\Delta H$, and $\nu$ has a frequency component, $\kappa$, a
temperature dependence, and an entropy component $\Delta S$
\cite{Gottstein:98}, 
\beq 
k = \kappa \frac{k_B T}{h} e^{\frac{\Delta S}{k_B}} e^{\frac{-\Delta H}{k_B
    T}},
\label{eq:Erying}
\eeq 
where, $h$ is Planck's constant. It follows that for the non-interacting case,
$\epsilon=0$, the frequency
component of $\nu$ is unchanged due to the fact that desorption and diffusion
events are unaffected by the presence of nearest neighbors,
and the changes in $\Delta S$ are perhaps sufficiently small in this regime, since the number 
of available microstates from one state to the next does not change significantly with the desorption
or diffusion of a single particle. In
Fig.~\ref{fig:nuvscov} we observe that $\nu$ always starts and ends at, or
close to, $1$ for all regimes: during the initial and final phases of the
desorption process, the entropy is at its lowest, and approaches zero at the
very beginning and end. At intermediate times/temperatures the entropy
increases due to the number of microstates that now become available.  In all
cases, $\epsilon \ge 0$, the initial phase of desorption occurs through eating
away of the large, percolating, connected cluster of occupied sites. Yet, in
the non-interacting case, there is a lack of correlation as to which sites
become unoccupied, while with increasing $\epsilon$, site occupation is
correlated over longer timescales as desorption due to the fact that
desorption and diffusion events are slowed down. These enhanced correlations
can be quantified through the time autocorrelation function for site
occupation: $C_{j} = <\sigma_{j}(t+\tau)\sigma_{j}(t)>$, where $\sigma$ is the
site occupation number, which takes on the values 0 or 1, and $\tau$ is the
time lag over which correlations are measured. The results of
Fig.~\ref{fig:Cit} (left) demonstrate that correlations in site occupation
become greatly enhanced as $\epsilon$ increases.

As desorption proceeds, the occupied lattice starts to break up into connected
(smaller) island clusters of occupied sites, this has been observed
experimentally \cite{Gunther:14}. The number of clusters reaches a
maximum value at a coverage value which both depend on interaction strength as
indicated by the data shown in Fig.~\ref{fig:Cit}(right). We can interpret
these features as follows: for a given value of the coverage, systems with
stronger interactions are likely to exist in larger but fewer clusters due to
enhanced site correlations that persist to longer times/temperatures. If the
number of clusters decreases the entropy is expected to decrease, which is
reflected in the value of $\nu$. The data of Fig.~\ref{fig:Cit} shows that the
coverage at which the difference in the number of clusters is largest
coincides with the greatest difference in the $\nu$ values, consistent with
Fig.~\ref{fig:nuvscov}. And although the numerical difference in the number of
clusters of Fig.~\ref{fig:Cit} (right) at first sight appears insignificant,
amounting to approximately a $10\%$ difference, this becomes significantly
magnified when evaluating the entropy through counting the number of
accessible microstates.

\subsection{False kinetic compensation effects in Thermal Desorption}
In this section we explore a reason for a a false compensation effect in thermal desorption,
which is demonstrated by
Miller et al. in \cite{Miller:87}. Starting from the slope of the Arrhenius plot, 
when all explicit dependencies are considered \cite{Miller:87,Nieskens:03}:
\beq 
\frac{d \ln{\left(\dot\theta/\theta\right)}}{d \left(1/T\right)} =
-\frac{E_a(\theta)}{k_B} + \frac{d \theta}{d
  \left(1/T\right)}\left(\frac{\partial{\ln{\nu (\theta)}}}{\partial{\theta}}
  -\frac{1}{k_B T}\frac{\partial{E_a (\theta)}}{\partial{\theta}}\right).
\label{eq:Cdepslope}
\eeq 
Here the second order terms contained within the parentheses in
produce a non-constant
slope and can only be ignored if: (1) the parameters are constant, (2)
if measurements are made over a region where the change in the coverage is vanishingly small
$\left(\frac{d \theta}{d
  \left(1/T\right)} \sim 0 \right)$, or (3) if
 the second order terms sum to zero, as a special manifestation of the compensation effect.
Assuming the latter, the following
differential equation results: $\frac{\partial{\ln{\nu (\theta)}}}{\partial{\theta}}
  =\frac{1}{k_B T}\frac{\partial{E_a (\theta)}}{\partial{\theta}}$, and its
solution reduces back to
Eq.~\ref{eq:KCE} \cite{Miller:87}.
This approach also suggests that the variations in the parameters
occur in the same direction, and with the same, or almost the same 
magnitude, thus the changes in $E_a$ and $\nu$ during the desorption 
process are ignored \cite{Nieskens:03}, and  it also forces a straight line
through the data that will likely lead to a false KCE \cite{Miller:87,Nieskens:03}.
Our results in Figs.~\ref{fig:nuvscov} and~\ref{fig:AplotsEnu}
do not support this assumption.
The increasing curvature of the Arrhenius plots in
Fig.~\ref{fig:Aplots} is the first visible indicator, 
if the variations in the parameters were effectively in the same direction and with the same
or almost the same
magnitude, then the plots, at the correct
order, should always yield straight lines.

\begin{figure}[h!]
    \includegraphics[width=0.23\textwidth]{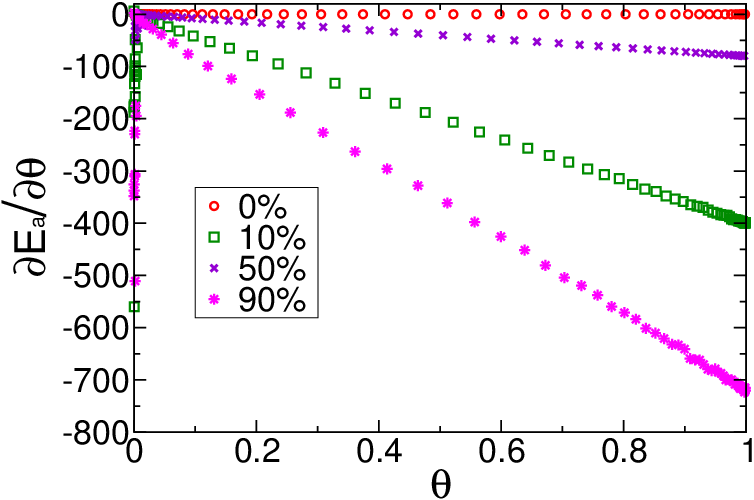}
    \includegraphics[width=0.23\textwidth]{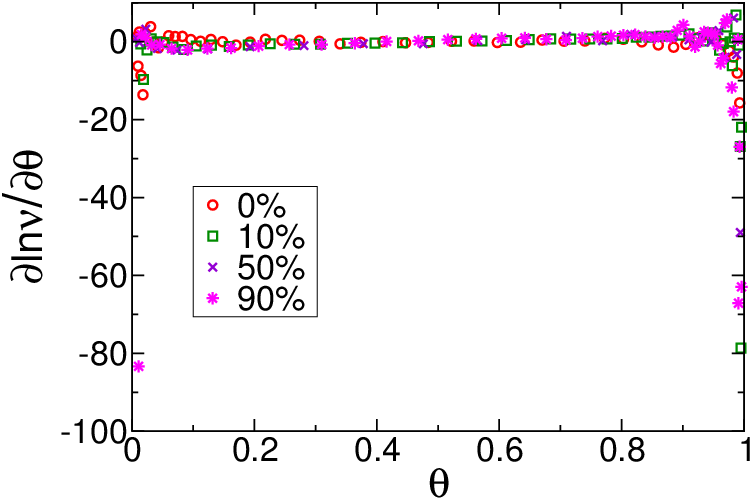}
    \caption{(Color Online) Second order, coverage-dependent terms appearing
      in the parantheses of Eq.~\ref{eq:Cdepslope}: Derivatives of, $E_{a}$
      (left panel) and $\ln{\nu}$ (right panel) with respect to coverage
      $\theta$.}
\label{fig:order2}
\end{figure}

To further emphasize this point we
calculated the second order terms 
in \ref{eq:Cdepslope} using our numerical data.
The results are shown in Fig.~\ref{fig:order2}.

\begin{figure}
    \includegraphics[width=0.45\textwidth]{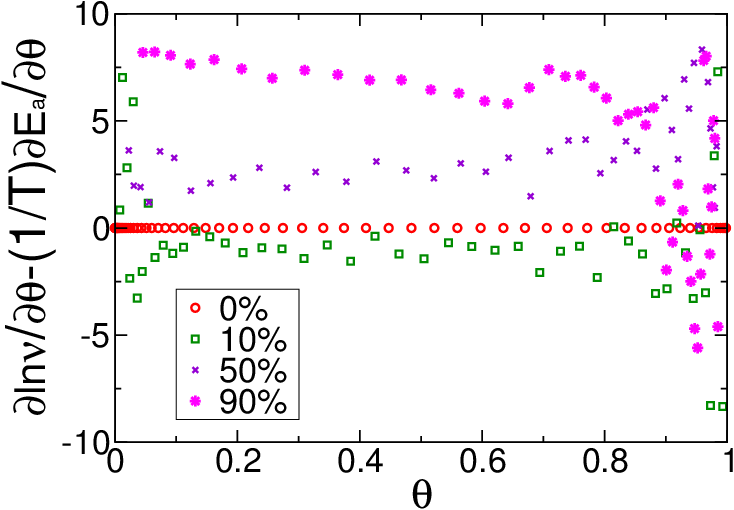}
    \caption{(Color Online) Second order terms in Eq.~\ref{eq:Cdepslope}
      $\frac{\partial{\ln{\nu (\theta)}}}{\partial{\theta}} -\frac{1}{k_B
        T}\frac{\partial{E_a (\theta)}}{\partial{\theta}} $, the terms only
      add to zero in the non-interacting regime, this implies that the second order
  terms in the slope of the Arrhenius plot can only be ignored in that instance.}
  \label{fig:order2trms}
\end{figure}

In Fig.~\ref{fig:order2} the derivative of $E_{a}$ exhibits much larger
variations (left panel) than $\ln{\nu}$ (right panel). When the terms are
added, according to Eq.~\ref{eq:Cdepslope}, the result is only zero in the
non-interacting case, as expected, otherwise it yields a non-zero, finite
contribution, which is plotted in
Fig.~\ref{fig:order2trms}.
It should be noted that the factor $\sfrac{1}{k_B T}$ attenuates variations in
$\frac{\partial{E_a (\theta)}}{\partial{\theta}}$, and, while these second
order terms are not particularly large, they yield a finite non-zero
contribution (see Fig.~\ref{fig:Aplots}), that, when ignored, forces a straight line 
through the data.

\subsection{Isokinetic relation (IKR)}

\begin{figure}
    \includegraphics[width=0.45\textwidth]{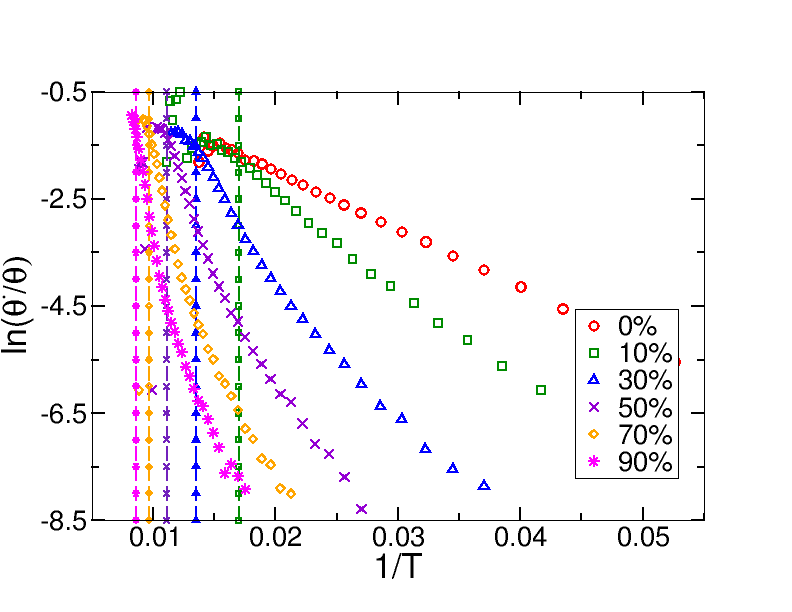}
    \caption{(Color Online) Approximate compensation temperatures
for all interaction regimes, indicated by the dashed lines in the same color
as the corresponding Arrhenius plots (symbols). The region where these compensation 
temperatures are observed correspond to low coverage and high temperature, where the
effects of lateral interactions on the rates of desorption $k$ start to become negligible.}
  \label{fig:IKR}
\end{figure}

In the context of isokinetic equilibrium, it can be seen in Fig.~\ref{fig:Aplots} (bottom panel) that the plots
exhibit a tendency
to converge towards the region of high temperature and low coverage. However, and as previously stated,
we do not observe that the plots cross at one value of $T_c$, but
all interacting curves seem to reach the non-interacting one, or an extrapolation
of it, at different temperatures. We indicate approximate values of $T_c$ for each interaction
regime in Fig.~\ref{fig:IKR}, and the numerical values of $E_a$, $\nu$ and $\ln{k}$ at those temperatures
are displayed in table~\ref{table1}.
\begin{table}[h]
\caption{Numerical values of the parameters, $E_{a}$ and $\nu$, at approximate
compensation temperatures, where the Arrhenius plots of 
the different interaction regimes reach the non-interacting plot, or an extrapolation of this curve.}
\label{table1}
\centering
  \begin{tabular}{ | c | c | c | c | r  |}
    \hline
    $\epsilon$ & Crossing Temp. & $E_a$ & $\nu$ & $\ln{k}$\\ \hline
    10$\%$ & 60 & -100.3474 & $1.00678 $ & $-1.67177 $\\ \hline
     30$\%$ & 78 & -100.118027 &  $ 0.9304 $ & $ -1.35568$\\ \hline 
     50$\%$ & 89 & -100.64 & $1.003921$ & $-1.1268736$\\ \hline 
     70$\%$ & 104 & -100.319222 & $0.91164$ & $-1.05712$ \\ \hline 
     90$\%$ & 118 & -100.235456 & $0.8753$ & $-0.98268$\\ \hline 
  \end{tabular}
\end{table}
As observed, the values of $E_a$ are very close
to that assigned to the binding energy of the
surface $E_b =100$. 
This information and the observed convergence towards
the region of low coverage and high temperature, implies
that this resembles an isokinetic relation.
We did not rely too heavily on the numerical values of the prefactor $\nu$, since
this parameter remains close to 
$1$ throughout the desorption process (see Fig.~\ref{fig:nuvscov}),
and the rates become close, but not exactly the same. Nevertheless, the
significance of this IKR is that \textit{it does not 
occur because of a compensation
between $E_a$ and $\ln{\nu}$}, but due to a 
transition
to the non-interacting regime, where
the rates are governed solely by the parameters of the
surface, and is therefore in agreement with the rates becoming
independent of external parameters and perturbations \cite{Freed:11}.

\subsection{Weak Adsorbate Interactions}
In this section we show numerical results for the regime of weak lateral
interactions, specifically for adsorbate-adsorbate interaction strengths $\epsilon \le
10\%$ of the binding energy. As mentioned before, some physisorption 
experimental studies fall in this category, such as the
desorption of Xe from a graphite surface \cite{Ulbricht:02}. In this regime
the Arrhenius plots are fairly close to linear, and estimation of the parameters
using the slope and intercept method of analysis
are more accurate to within a few percent of the actual calculations (see Table
\ref{table2}). 
\begin{figure}
    \includegraphics[width=0.48\textwidth]{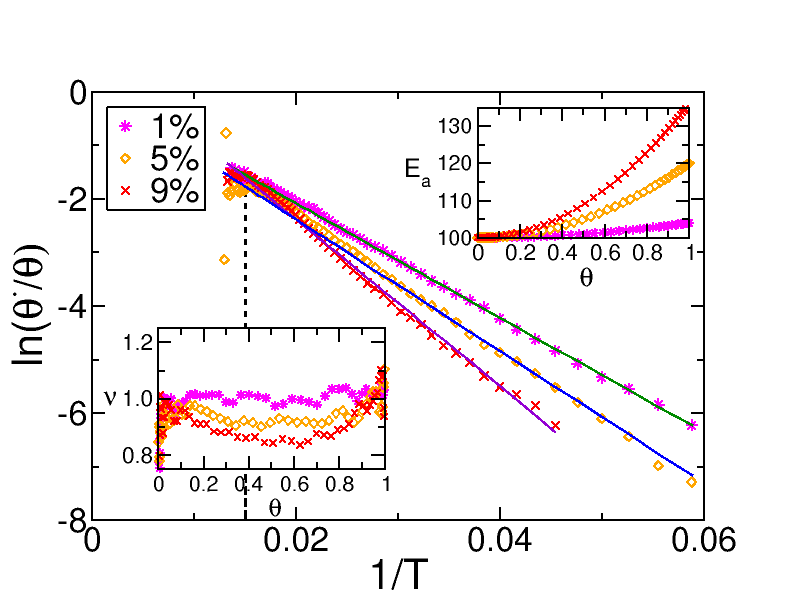}
    \caption{(Color Online) Arrhenius plots for interaction strengths, $1\%$,
      $5\%$ and $9\%$ of the surface binding energy. Insets show the
      coverage-dependence of the activation energy, $E_{a}$, and
      preexponential factor, $\nu$. Symbols are simulation data and solid
      lines are best linear fits to the data. Comparing between the linear
      fitting and exact calculations of the Arrhenius parameters yields the
      following error estimates in \{$E_{a},\nu$\}: \{2.9\%,4\%\},
      \{2.5\%,13\%\}, and \{7.5\%,123\%\}, with increasing interaction
      strength. The vertical dashed line indicates a possible compensation temperature.}
\label{fig:aplots}
\end{figure}
\begin{table}[h]
\caption{Arrhenius parameters, $E_{a}$ and $\nu$, obtained using a forced linear fit to the
  Arrhenius plots compared with the directly calculated values from the simulations.}
\label{table2}
\centering
  \begin{tabular}{ | c | c | c | c | r  |}
    \hline
    $\epsilon$ & $E_a$ max. & $E_a$ linear fit & $<\nu>$ calculated & $\nu$ linear fit  \\ \hline
    0$\%$ & 100 & 102 & 0.9899 & 1.0 \\ \hline
     1$\%$ & 104 & 107 &  $ 1.019 $ &1.06  \\ \hline 
     5$\%$ & 120 & 123 & $0.95688$ & 1.08  \\ \hline 
     9$\%$ & 145 & 156 & $0.931$ & 2.08  \\ \hline 
     10$\%$ & 140 & 167 & $0.931$ & 2.4  \\ \hline 
     30$\%$ & 220 & 292 & $0.89361$ & 8.18  \\ \hline
     90$\%$ & 460 & 785 & $0.88759$ &   121.94 \\ \hline
  \end{tabular}
\end{table}
In table~\ref{table2} we tabulate the values of the parameters, $E_a$ and $\nu$ extracted by
using traditional methods of analysis vs. our numerically calculated results
over our complete range of adsorbate-adsorbate interaction strengths
$\epsilon$.  Specifically, we compare how a linear fitting of the data leads
to increasing differences from the exact calculated values. While for weaker
interaction energies, the parameters from a linear fit procedure provides
reasonably accurate estimates, at stronger interactions deviations grow
dramatically.

These results demonstrate how `unusually high' preexponential factors can be obtained
from extracting the parameters through enforcing the 
traditional linear fitting methods in the regime of strong
interactions. The large magnitude of the extracted values of $\nu$ in the
regimes of $30\%$ and $90\%$ interaction 
strenght have been
attributed to a strong coverage dependence \cite{Zhdanov:89,Zhdanov:81,Seebauer:88,Pfnur:78},
 but here we have
demonstrated that this is not the case and that it is likely that this issue
might be a consequence of forced linearization.

\begin{figure}
    \includegraphics[width=0.48\textwidth]{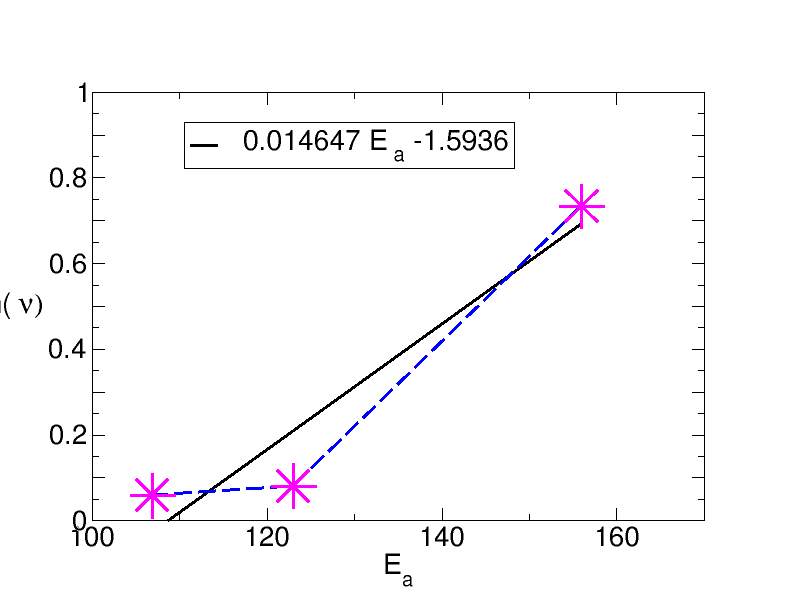}
    \caption{(Color Online) Constable plots using fitted parameters $E_a$ and 
    $\ln{\nu}$ for interaction strength regimes of $1\%$,
      $5\%$ and $9\%$ of the surface binding energy.
      A linear fit to this plot yields a slope of $0.014647$, which
	corresponds to temperature $T = 68.27$, and $y-$intercept 
	$-1.5936$.}
\label{fig:Constable1}
\end{figure}

In the context of the IKR, we can see the Arrhenius plots in Fig.~\ref{fig:aplots} 
overlapping in the 
region of low coverage and high temperature, towards the end of the 
desorption process.
Using the Arrhenius parameters in table~\ref{table2}, we constructed
the Constable plot of $\ln{\nu}$ 
vs. $E_a$ in Fig.~\ref{fig:Constable1}, for the regimes with $\epsilon = 1\%$, $5\%$
and $9\%$ interaction strength. 
A linear fit to this Constable plot yields a slope with $T_c = 68.27$
and $\ln{k} = -1.5936$, with correlation coefficient of $0.9554181$. Using
$T_c \approx 68$, we extracted the values of $E_a$, $\nu$ and $\ln{k}$ at that point.
The results are displayed on 
table~\ref{table3}, where the numerical values of $E_a$ deviate by a small amount from 
the surface binding energy $E_b = 100$, therefore we can consider this to be 
the point of transition to the non-interacting regime. This point is indicated in
Fig.~\ref{fig:aplots}. At $T_c = 68$ the Arrhenius plots begin to overlap, and we should point out that
henceforth, we can find higher temperature values for which the corresponding values of $E_a$, $\ln{\nu}$ and
$\ln{k}$ also fall within the non-interacting regime.

The numerically calculated values of $\nu$ at $T_c = 68$
are also close to unity, with some variations. However since the prefactor stays relatively close to 
$1$ throughout the process, we once again rely more strongly on
the numerical values of $E_a$ to make this assessment. The values of 
$\ln{k}$ are also displayed on table~\ref{table3}, and, as can be seen these
values are closer to the intercept of the Constable plot.
We previously mentioned that for a linear Constable plot to predict
an IKR, it is necessary that the linear correlation coefficient is $1$. Here we 
obtained a strong linear correlation and our analysis yields a compensation
temperature in the region where the Arrhenius plots acquire close values of the rate.
This demonstrates that, within a regime where molecular interactions are fairly weak compared
to the energy of activation, the 
prediction of the KCE and IKR can be performed using the Constable plot.
Nevertheless, the variations in the parameters exhibit the same behavior we observed
with stronger interactions. And this implies that compensation effects can occur in non-Arrhenius curves,
but cannot be determined using the methods outlined here.
\begin{table}[h]
\caption{Numerical values of the parameters, $E_{a}$, $\nu$ and $\ln{k}$, at
$T = 69$, obtained from the linear fit to the Constable plot in 
Fig.~\ref{fig:Constable1}. The intercept yields a value of $-1.5936$ in the 
same linear fit.}
\label{table3}
\centering
  \begin{tabular}{ | c | c | c | c | r  |}
    \hline
    $\epsilon$ & $T_c$ & $E_a$ & $\nu$ & $\ln{k}$\\ \hline
    1$\%$ & 69 & -100.000169 & $1.124706 $ & $-1.58811$\\ \hline
     5$\%$ & 69 & -100.002511 &  $ 0.879276 $ & $ -1.59928$\\ \hline 
     9$\%$ & 69 & -100.01026 & $1.021269$ & $-1.44969$\\ \hline 
  \end{tabular}
\end{table}
\section{Conclusion}
We have demonstrated that the Arrhenius parameters effectively 
vary in a manner consistent with a kinetic 
compensation effect.
However the variations in the prefactor
are not large enough to completely offset those in 
$E_a$, and thus a net partial compensation occurs instead.
As the interaction strength increases, so does the peak temperature,
which is expected, but it also means that the variations in $\nu$ are not 
sufficient to keep the rates almost unchanged. Although it could be 
argued that this is due to the relatively large difference
between regimes of interaction strenght (here it is $10\%$), we observed this
same behavior for desorption curves with smaller differences.

A Constable plot in the regime of weak interactions ($< 10\%$) yields a slope
that corresponds to a temperature where the rates acquire close values, but not the
exact same, and
the Arrhenius 
plots begin to overlap. This is consistent with the rates becoming independent
of external perturbations. We confirmed this by showing that, at this point, the 
values of $E_a$ and $\ln{\nu}$ become very close to the bare parameters of the
surface.
We observed a similar convergence in the stronger interaction regimes
($\ge 10\%$ of the binding energy), however the
temperatures at which the rates become independent of
adsorbate interactions are considerably different for each Arrhenius plot in this
case.
Nevertheless, in both instances,
the plots tend to overlap \textit{because 
of a transition to a regime where lateral interactions become
negligible}, and not due to mutual compensation.

The data points on the Constable plot represent the \textit{apparent} Arrhenius 
parameters, and although this method of analysis works well within the
weak interaction regime, our stronger interaction regime data shows that 
there can be instances of the KCE in instances where
non-Arrhenius behavior is observed, and that these 
effects may be overlooked due to the strict criteria used to determine the occurrence a KCE
and the IKR.

The behavior we observe could very likely help elucidate and compensation effects
and isokinetic relations
observed in other systems. A better understanding of this phenomenon can help achieve
controlled activated events and provide a means to accurately
parameterize many biological, chemical and physical processes that share
common features in their compensation effects.

\begin{acknowledgements} 
N. Zuniga-Hansen and M. M. Calbi acknowledge support provided by the National
Science Foundation through grant CBET-1064384. 

The authors would also like to thank Professor Phillip Sprunger for valuable discussions.
\end{acknowledgements} 




\end{document}